\documentclass[aps,superscriptaddress,showpacs,twoside,notitlepage,prl]{revtex4-1}
\usepackage{t1enc}
\usepackage[english]{babel}
\usepackage[utf8]{inputenc}
\usepackage{graphicx}
\usepackage{psfrag}
\usepackage{amsmath}
\usepackage{amssymb}
\usepackage{bbold}
\usepackage{hyperref}

\def\d{\mathrm{d}}
\def\e{\mathrm{e}}

\def\lag{{\mathcal{L}}}

\def\kihagy#1{}

\begin{document}
\title{Plane waves as tractor beams}
\author{P\'eter Forg\'acs}
\affiliation{MTA Wigner RCP RMI, H1525 Budapest, POB 49}
\affiliation{LMPT CNRS UMR7350, Université de Tours, Parc de Grandmont, 37200 Tours, France}
\author{\'Arp\'ad Luk\'acs}
\affiliation{MTA Wigner RCP RMI, H1525 Budapest, POB 49}
\author{Tomasz Roma\'nczukiewicz}
\affiliation{Institute of Physics, Jagiellonian University, Reymonta 4, 30-059 Cracow, Poland}

\begin{abstract}
It is shown that in a large class of systems plane waves can act as {\sl tractor beams}: i.e.,\ an incident plane wave can exert a {\sl pulling}
force on the scatterer. 
The underlying physical mechanism for the pulling force is
due to the sufficiently strong scattering of the incoming wave into another mode having a larger wave number, in which case {\sl excess momentum} is created {\sl behind} the scatterer.
Such a tractor beam or {\sl negative radiation pressure} effect arises naturally in systems where the coupling between the scattering channels is due to Aharonov--Bohm (AB)
gauge potentials. It is demonstrated that this effect is also present if the AB potential is an induced, (``artificial'') gauge potential such as the one found in J.~March-Russell, J.~Preskill, F.~Wilczek, {\sl Phys.\ Rev.\ Lett.} {\bf 58} 2567 (1992).
\end{abstract}
\maketitle

\paragraph{Introduction}
It seems to be a hitherto unremarked phenomenon, that in multi-channel scatterings the force acting on the scatterer can be counter-intuitive. We show
that for a large class of scattering systems with at least two channels with different dispersion relations, an incoming plane wave can exert a pulling force
on the scatterer (tractor beam effect or negative radiation pressure, NRP).
Models where negative radiation pressure appears have many physical applications, ranging from various condensed matter systems with vortices
to cosmic strings.

We shall consider in more detail systems with vorticity, where the coupling between the two channels is due to an Aharonov--Bohm (AB) potential.
Typical cases of such systems are vortices, where the AB potential is induced by the ``internal frame dragging'' due to the breaking of a global U(1) symmetry \cite{W}. For a recent review on systems with similar
``artificial gauge potential'' we refer to the review \cite{art}.
Another noteworthy example is the scattering of sound on a superfluid vortex \cite{vort,ior} or the scattering of various fields on cosmic strings \cite{W,rel,FLRv}.
Our prototype example will be the model studied in Ref.\ \cite{W}, which describes the
scattering of scalar fields on a global vortex or that of neutral particles on a magnetic vortex. Unlike Ref.\ \cite{W}, we consider the case when both
channels are open. We show, that for a large range
of the parameters of the model, the incident plane wave acts as a tractor beam.

Negative radiation pressure on localized solutions (kinks) due to an incoming plane wave has already been observed in one dimensional
classical field theories \cite{FLR, trom}. In that case NRP is induced in systems with a small reflection coefficient and sufficiently strong
conversion to higher harmonics by the nonlinearities \cite{FLR}. More recently, there has been an upsurge of interest for
tractor beams in optics, with light beams specifically prepared for the scatterer in a way that it is scattered dominantly in the direction behind the scatterer \cite{sukhov}.

\paragraph{Setup} In this paper, we consider stationary scattering, in one and two dimensional cases. The incoming plane wave propagates in
the positive direction along the $x$ axis, with the scatterer placed at the origin.
We assume two channels $u$, $d$, with wave numbers $k_u$, $k_d$,
related to the wave frequency $\omega$, by the dispersion relations $k_i^2 + m_i^2 = \omega^2$.

\paragraph{One dimensional case}
For the case of an incoming wave in the $u$ channel with amplitude $A_u$ and wave number $k_u$,
the force acting on the scatterer follows from momentum conservation:
\begin{equation}
  \label{eq:1dforce}
  F_u/A_u^2 = k_u(1+|R_{uu}|^2-|T_{uu}|^2) + k_d(|R_{du}|^2-|T_{du}|^2)\,,
\end{equation}
where $R_{ij}$ resp.\ $T_{ij}$ denote the transmission resp.\ reflection coefficients
for an incoming wave of type $j$ into channel $i$,
and $k_d$ is the wave number in the $d$ channel.
Eq.\ (\ref{eq:1dforce}) follows immediately by taking into account that
a plane wave $A\exp(ikx)/\sqrt{k}$, carries momentum $k |A|^2$.
We note that the energy flux of such a wave is $\omega |A|^2$.

The conservation of energy implies that the energy flux of the outgoing waves is equal to that of the incoming wave, therefore
\begin{equation}
  \label{eq:encons}
  \sum_i(|R_{ij}|^2 + |T_{ij}|^2)=1\,,
\end{equation}
which corresponds to the unitarity of the S-matrix. By combining eqns.\ (\ref{eq:1dforce}) and (\ref{eq:encons}), the first term in
eq.\ (\ref{eq:1dforce}) is positive, and thus for a single channel scattering Eq.\eqref{eq:1dforce} reduces to the familiar radiation pressure formula $F = 2k|R|^2 > 0$.

Already when the coupling of the channels is due to a simple step potential, the forward scattering amplitude from channel $u$ to $d$, $T_{du}$, can become
large enough to reverse the direction of the force (\ref{eq:1dforce}) acting on the scatterer, provided that $k_d > k_u$, (which is the case when, e.g.,\ $m_u > m_d$), that is
NRP is induced by flavor changing scattering.

Physically, such a scenario can be realized by the low energy scattering of neutral particles with magnetic moment,
in a magnetic field perpendicular to their propagation (e.g.,\ $B_z <  0$, the two species being spin up/down).
The scatterer is a region with magnetic field along the $x$ axis.

We have illustrated that NRP on a scatterer appears for various one dimensional systems. The physical mechanism for the NRP in this simple setting is
that if an incoming wave in the $u$ channel is scattered sufficiently strongly to the $d$ channel with $k_d > k_u$,
excess momentum is created behind the scatterer, since
the scattered wave in the $d$ (``light'') channel carries more momentum than the incoming one.

\paragraph{Force acting on the scatterer in planar problems}
As we shall show, NRP occurs for multi-channel scattering in a number of 2 dimensional systems,
in spite of waves being scattered in all directions, not just forward and backwards as in 1 dimension.
Exploiting the conservation equations $\dot{\mathcal P} = -\nabla {\bf T}$ where
${\mathcal{P}}$ is the momentum density of the field, we obtain the force acting on the scatterer in two dimensions:
\begin{equation}
  \label{eq:force}
  {\bf F} = -\lim_{R\to\infty}\int_{-\pi}^\pi \bar{\bf T}{\bf e}_r R\d\vartheta\,,
\end{equation}
where ${\bf T}$ is the stress tensor, the overbar denotes time averaging,
and $r,\vartheta$ denote the polar coordinates of the plane. To compute ${\bf T}$ asymptotically, we assume that in the far field region the waves can be considered freely propagating.
Assuming $n$ channels, for each partial wave $\ell$, the $S$-matrix elements form an $n\times n$ unitary matrix $S_\ell$. Evaluating the expression (\ref{eq:force}) we
obtain a master equation for the force ${\bf F}$, in terms of $S_\ell$:
\begin{equation}
  \label{eq:fpw}
  {\bf F} = F_x+i F_y=-4 \sum_\ell \left\{ A^\dagger S_{\ell+1}^\dagger K S_\ell A
    - A^\dagger  K  A
\right\}\,,
\end{equation}
where $A=(A_1,\dots,A_n)^T$, $K=\text{diag}(k_1,\dots,k_n)$ and $\dagger$ denotes the adjoint (transposed conjugate).
Eq.\ (\ref{eq:fpw}) can also derived for potential scattering in quantum mechanics (in $\hbar=1$ units).

Analogously to the one dimensional case, it is not difficult to see that for a free, two component
wave $(u,d)$,  coupled in a disk-like region $r<L$, by a constant potential, NRP is observed for various masses of the particles in a large
range of frequencies.
We should like to mention that the force acting on the solenoid for
the Aharonov--Bohm (AB) scattering \cite{AB}, follows from our general result Eq.\ (\ref{eq:fpw}). To show this 
we recall that the scattering phases in the partial wave expansion for the AB problem are given as $S_\ell = \exp(2i\delta_\ell)$,
$\delta_\ell = \pi(\ell-\nu)/2$, with $\nu^2=(\ell-\Phi)^2$, where $\Phi$ is the magnetic flux in the solenoid.
Then from Eq.\ (\ref{eq:fpw}) the well known result for the longitudinal and transversal component of the force \cite{ABf} follows:
\begin{equation}\label{eq:ABforce}
F_x+iF_y = -4|A|^2k(\e^{-2i\pi\Phi}-1)\,.
\end{equation}
It is a somewhat striking result that the force acting on the solenoid has also a transversal component.
As it is clear by now this is due to the breaking of the reflection symmetry with respect to the $x$ axis by the magnetic flux (direction of circulation).

The result \eqref{eq:ABforce} has other applications as well.
For example the force exerted on a superfluid vortex by the scattering of sound waves in the Gross-Pitaevskii (GP) model
is given by Eq.\ \eqref{eq:ABforce} with a suitable replacement
of the flux $\Phi \propto \omega$ \cite{vort}.
The longitudinal component, $F_x$, is the acoustic drag, and the transversal component is known as the Iordanskii force \cite{ior}.

\paragraph{Scattering on a vortex}
In Ref.\ \cite{W} it has been shown that in the low momentum transfer limit, the scattering cross section of a complex scalar $\psi$, on a vortex in a GP-type model
is very close to that of the AB problem.
In Ref.\ \cite{W} the problem is reduced to a single channel scattering, which is valid below the threshold defined by the larger mass eigenvalue $m_u$.
In the following we shall show that at higher energies when both channels are open, the scattering of the $u$ mode induces NRP on the vortex.
The coupling term between $\psi$ and the symmetry breaking complex scalar field $\phi$, is given by $\Delta\lag = g\phi \psi^2 + \text{c.c.}$.
The field equations for the field $\psi$ are
\begin{equation}
  \label{eq:Weom}
  (\omega^2 -m^2 +\nabla^2) \psi = 2 g \phi^* \psi^*\,.
\end{equation}
Far from the vortex $\phi\sim v \exp(i\vartheta)$, therefore the mass eigenstates of the
potential are defined by the transformation $(\psi,\psi^*)^T = \sqrt{2} U \rho$, $\rho=(u,d)^T$.
$U$ is a unitary matrix given as
\begin{equation}
  \label{eq:unittrf}
  \sqrt{2} U = \begin{pmatrix} \e^{-i\vartheta/2} & i \e^{-i\vartheta/2} \\ \e^{i\vartheta/2} & -i\e^{i\vartheta/2} \end{pmatrix}\,.
\end{equation}
Transformation \eqref{eq:unittrf} yields a heavy mode $u$ ($m_u^2= m^2+2 g v$) and a light mode $d$ ($m_d^2=m^2-2 g v$). Eqs.\ \eqref{eq:Weom} can be written as
\begin{equation}
  \label{eq:meig}
  \left(\nabla + i {\bf A} \frac{\sigma_2}{2}\right)^2 \rho - K^2 \rho =0\,, \quad \sigma_2=\begin{pmatrix} & -i \\ i & \end{pmatrix}\,,
\end{equation}
where ${\bf A}={\bf e}_\vartheta/r$ is an induced, ``artificial'' gauge potential with vorticity (or winding number) $\pm1/2$, ${\bf e}_\vartheta$ is the angular unit vector.
Note, that the $\exp(i\theta/2)$ terms in Eq.\ \eqref{eq:unittrf} induce anti-periodic boundary conditions on $\rho$.

Partial waves are introduced taking into account the vortex number of ${\bf A}$, as
\begin{equation}
  \label{eq:pwaves}
  (u,d)=\sum_{\ell=-\infty}^\infty \e^{i(\ell+\gamma)\vartheta}(u_\ell(r), d_\ell(r))\,,
\end{equation}
with $\gamma=1/2$. The radial functions satisfy
\begin{equation}
  \label{eq:radeq}
  \begin{aligned}
    u_\ell''+\frac{u_\ell'}{r}-\frac{{\eta}_u^2}{r^2}u_\ell +\frac{c}{r^2}d_\ell + k_u^2 u_\ell &= 0\,, \\
    d_\ell''+\frac{d_\ell'}{r}-\frac{{\eta}_d^2}{r^2}d_\ell +\frac{{c}^*}{r^2}u_\ell + k_d^2 d_\ell &= 0\,,
  \end{aligned}
\end{equation}
where ${\eta}_u^2={\eta}_d^2=(\ell+1/2)^2+1/4$ and ${c}=i(\ell+1/2)$. We use rescaled variables such that
$v=1$, $m_u=2$ (threshold at $\omega=2$), and for simplicity sake, present numerical data for $m_d=1$.

We have computed the $S$-matrix elements by solving numerically Eqs.\ \eqref{eq:radeq}
for $0\leq\ell\leq12$ for various values of the frequency $2<\omega\leq4$ and evaluated the force acting on the (point)vortex from the master Eq.\ \eqref{eq:fpw}, comp.\ Figs. 1-3.
(We have verified that in this frequency range partial waves with $\ell\geq12$ give negligible contribution.)
To understand the results for the coupled channel scattering better,
we present some analytical results based on perturbation theory
in the coupling, $c$, near the threshold of the $u$-channel ($k_u\ll1$).
We start at the threshold ($k_u=0$) when channel $u$ is still closed while channel $d$ is open.
Then the $0$-th order solutions are: $u^{(0)}_\ell=0$; $d_\ell^{(0)} = i^\ell \e^{i{\xi}_d} J_{{\eta}_d}(k_d r)$, with $\xi_d = (\ell-{\eta}_d)\pi/2$.
Furthermore not only the first order correction $u^{(1)}_\ell$, can be calculated in closed form,
($d_\ell^{(1)}=0$) but also the asymptotic form of the second order one, $d_\ell^{(2)}$.
The $S$-matrix is reduced to a single scattering phase $\delta_d$, given as:
\begin{equation}
  \label{eq:nu2}
  \delta_d={\xi}_d+\Delta\,,\quad{\rm with\ }\Delta=
 \arctan\left[ \frac{|{c}|^2 \pi/4}{{\eta}_u^2 {\eta}_d+ {\eta}_u {\eta}_d^2}\right]\,.
\end{equation}
What one learns from the above is that $\delta_d$ at the threshold is determined not just by the expected AB phase shift corresponding to $\Phi=1/2$ but that
the 2nd order correction, $\Delta$, is still important for not too large values of $\ell$.
The perturbative computation of the $S$-matrix above the threshold is significantly simplified by taking into account the above phase correction, Eq.\ (\ref{eq:nu2}),
for the $0$-th order $d$ channel wave, i.e.,\
$d_\ell^{(0)} = i^\ell \e^{i{\delta}_d} J_{{\nu}_d}(k_d r)$, where $\nu_d = \ell-2\delta_d/\pi$. Then it is sufficient to go to just 1st order in perturbation theory
to obtain the leading part of the $S$-matrix close to the threshold. We find
\begin{equation}
  \label{eq:Smat}
  S_\ell \approx \frac{1}{N}\begin{pmatrix}
    \e^{2 i\delta_u} & \e^{i(\delta_u + \delta_d)} i\pi {c}I \\
    \e^{i(\delta_u + \delta_d)} i\pi {c}^*I & \e^{2 i\delta_d}
\end{pmatrix}\,,
\end{equation}
with $I$ given as
\begin{equation}
  \label{eq:intF}
 I=\left( \frac{k_u}{k_d} \right)^{\nu_u} \Gamma(\bar{\nu})
 \frac{{}_2 F_1\left(\nu',\bar{\nu},\nu_u+1,k_u^2/k_d^2 \right)}
 {2\Gamma\left(1-\nu'\right)\Gamma(\nu_u+1)}\,,
\end{equation}
where $\nu_u = \eta_u$, $\delta_u=\pi(\ell-\nu_u)/2$), $\bar{\nu}=(\nu_u+\nu_d)/2$, $\nu'=(\nu_u-\nu_d)/2$, and $N$ is a normalization factor to achieve exact unitarity in each order.
Our perturbative results provide a satisfactory qualitative description
of the numerical solution of Eqs.\ \eqref{eq:radeq}, comp. Figs.\ 1-3.
We remark also that this perturbative computation induces relatively large errors
for small values of $\ell$, caused by the behavior at $r=0$ of the phase corrected $d$-channel wave.

\begin{figure}[h!]
{\psfrag{l}{\small$\ell$}
\noindent\hfil\includegraphics[scale=.33,angle=-90]{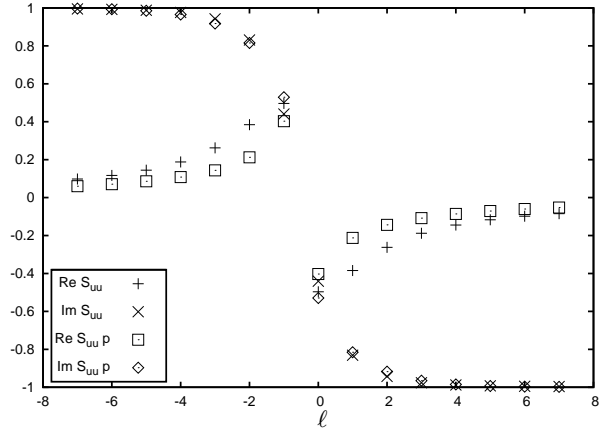}
}
\caption{Real and imaginary parts of $S_{uu}$ for $\omega=2.22$ in function of $\ell$,
computed numerically and perturbatively (p).
}
\label{fig:peco}
\end{figure}

\begin{figure}[h!]
\noindent\hfil\includegraphics[scale=.33,angle=-90]{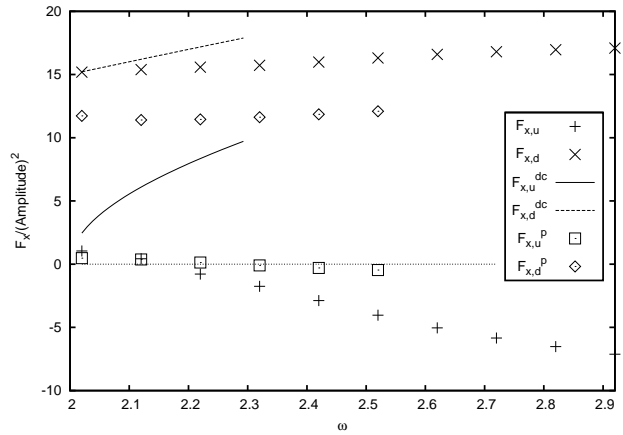}
\caption{The longitudinal force component $F_x$ as a function of the frequency $\omega$: the numerical result is compared to
the decoupled, (dc), resp.\ to the perturbative, (p), approximation. (dc) resp. (p) are depicted for $2<\omega\leq2.3$ resp.\ $2<\omega\leq2.5$.
For the incoming heavy mode $u$, the radiation pressure becomes negative at $\omega\approx 2.1557$.}
\label{fig:fpvx}
\end{figure}

\begin{figure}[h!]
\noindent\hfil\includegraphics[scale=.33,angle=-90]{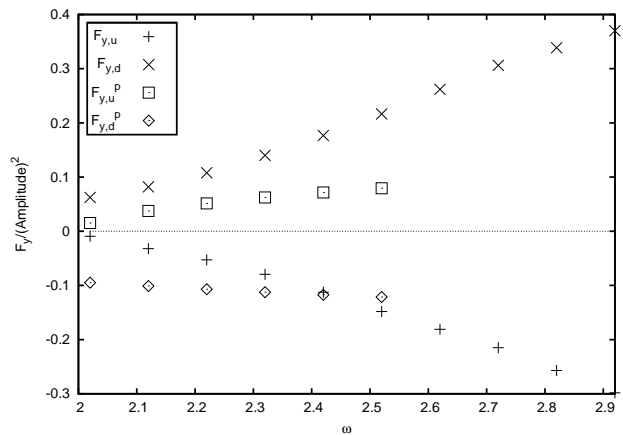}
\caption{The numerical and the perturbative (p) result for the transversal force component $F_y$ as function of the frequency $\omega$. (p) is depicted only for $2<\omega\lessapprox2.5$).}
\label{fig:fpvy}
\end{figure}

\begin{figure}[h!]
\noindent\hfil\includegraphics[scale=.33,angle=-90]{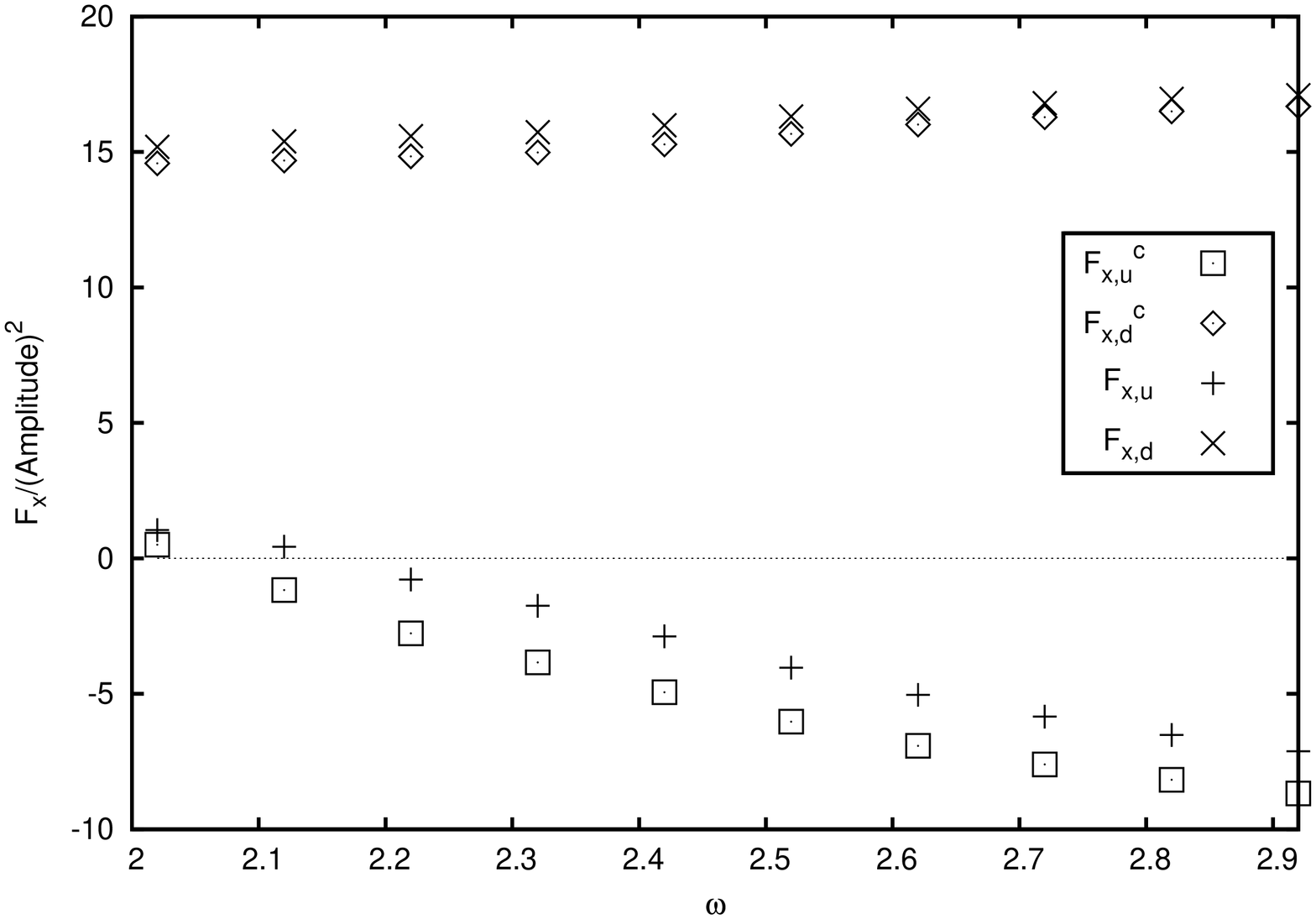}
\caption{The effect of the vortex core: the longitudinal force $F_x$ acting on a point vortex and on a vortex with a linear core (marked with `c'). The existence of a vortex
core enhances NRP.}
\label{fig:forcexcore}
\end{figure}

\paragraph{Force acting on the scatterer} Once the $S$-matrix is known, our master formula
Eq.\ (\ref{eq:fpw}) yields the force acting on the scatterer.
Consider first the decoupled problem, when the $S$-matrix elements are very simple, $S_\ell = \text{diag}(\e^{2i{\xi}_u}, \e^{2i{\xi}_d})$,
where ${\xi}_{u,d}=\pi(\ell-{\eta}_{u,d})$.
For the longitudinal component one finds: $F_{x,i} \approx 8.67 k_i$ (there is no induced NRP) and $F_y=0$ (there is no transversal force), comp. Fig.\ \ref{fig:fpvx}.
Note that the drag force due to the incoming $d$ mode is quite well reproduced in the decoupled approximation, while it is completely wrong for the $u$-mode.

Slightly above the threshold, the scattering from channel $u$ to $d$
(the $S_{du}$ term) can lead to NRP on the vortex, comp. Fig.\ 2.
The transversal force acting on the scatterer, $F_y$, is an analog of the Iordanskii force acting on vortices in superfluids \cite{vort,ior}.
In our case the total transversal force is entirely due to the coupling between the channels, comp. Fig.\ \ref{fig:fpvy}.

As Figs.\ \ref{fig:fpvx}, \ref{fig:fpvy}.\ illustrate, near threshold perturbation theory describes qualitatively the effects of the coupling between the two modes, losing
its validity for frequencies $\omega \gtrsim 2.4$. The main source of errors in our perturbative approach appears in the phases of the matrix elements of $S_{ud}$ and  $S_{du}$.

In the preceding analysis, the force acting on a scatterer was obtained from a stationary waveform. A condition for the validity of this description is that the outgoing
wave packets in both modes reach the asymptotic region (where the wave field can be described by the $S$-matrix) within a characteristic time scale, which is problematic
precisely at the threshold. For this reason, it is important for the validity of our perturbative approach  that the perturbative results are valid for energies away from the threshold.

\paragraph{Effect of a vortex profile}
Up to now we have considered point vortices, when the field configuration is approximated everywhere by its asymptotic form $\phi=v\exp(i\vartheta)$.
A more realistic vortex configuration, $\phi=f(r)\exp(i\vartheta)$, has of course a non-singular core region which can be described by a profile function $f$, satisfying
$f(0)=0$ and $f(r\to\infty)=v$. To illustrate how a vortex core affects our results we consider a simple profile function
$f(r\le R_c) = v r/R_c$, $f(r\ge R_c)=v$.
In Fig.\ \ref{fig:forcexcore}, the effect of a linear core on the longitudinal force is depicted, for $R_c=1$. It can be seen that the NRP effect is enhanced by the profile.
The reason for this enhancement can be understood in that a localized potential
affects significantly partial waves with small values of $\ell$, and brings the $\ell=0$ component of $\delta_u$ closer to
that of $\ell=1$, resulting in the reduction of the $\ell=-1,0$ force components.

\paragraph{Generalizations} A remarkably large class of scattering problems, like the perturbations of superfluid vortices \cite{vort}, perturbations
of global cosmic strings \cite{rel,FLRv} or the scattering of neutral spin-1/2 particles on a magnetic vortex \cite{W}
can be brought to the form of a multi-component AB-type scattering, with a potential matrix
\begin{equation}
  \label{eq:potmat}
 P= \begin{pmatrix} \mu_1^2 & \alpha\e^{in\vartheta} \\ \alpha^* \e^{-in\vartheta} & \mu_2^2\end{pmatrix}\,.
\end{equation}
Diagonalizing $P$ with a unitary matrix $U$ and introducing mass eigenstates with $\psi_i=\sqrt{2} U_{ij}\rho_j$,
we assume field equations of the form
\begin{equation}
  \label{eq:masseig}
  (\nabla - i T {\bf A})^2 \rho + K^2 \rho = 0\,,\quad
  T=\begin{pmatrix}e_1 & q \\ q^* & e_2 \end{pmatrix}\,.
\end{equation}
When partial waves are introduced, the coefficients of the $1/r^2$ term are given as ${\eta}_i^2=(\ell-\gamma-e_i)^2+|q|^2$ and $c=q(2(\ell+\gamma)-(e_1+e_2))$.
From this point on the perturbative
calculations can be carried out in the same way as for Eqs.\ \eqref{eq:radeq}.

A number of important problems can be considered in the framework of \eqref{eq:masseig}. They include the example considered in Ref.\ \cite{W}, the scattering of
neutral spin-1/2 particles on a magnetic vortex, when $q=i/2$, $k_d^2-k_u^2=gM$,
$M$ being the magnetization. Another noteworthy case is the scattering of
perturbations on superfluid vortices in the GP model with 1st order dynamics \cite{vort}, when $e_1=-e_2=-\omega/2+\dots$, $q=1$, $k_u^2=2-\sqrt{4+\omega^2}$, $k_d^2=2+\sqrt{4+\omega^2}$,
(note that the $u$ channel is closed for any value of $\omega$).
Vortex solutions in the GP model with 2nd order dynamics are used to model (straight) global cosmic strings, and scattering of various fields on them has been intensively studied \cite{rel}.
Scattering of scalar perturbations on cosmic strings fits naturally in our framework too,
$e_1=e_2=0$, $q=-1$, $k_u^2=\omega^2-4$, $k_d=\omega$.
We have carried out a detailed study of this problem  with the result
that by the scattering of the $u$-mode NRP appears \cite{FLRv}.
In this case the effect of the vortex core is crucial for the NRP, which is absent
in the point vortex limit. We also note that
for the global cosmic string, due to the symmetry of the radial equations, $\ell\to-\ell$, the transversal force is absent, $F_y=0$.
We have been able to confirm the validity of our computations by carrying out time-dependent numerical simulations of the field equations \cite{FLRv}.

\paragraph{Acknowledgements} This work has been supported by OTKA grant no.\ K101709.
\def\refttl#1{#1, }


\begin{thebibliography}{999}
\bibitem{W} J.~March-Russell, J.~Preskill, and F.~Wilczek, \refttl{Internal frame dragging and a global analog of the Aharonov--Bohm effect}
{\sl Phys.\ Rev.\ Letters} {\bf 58}, 2567 (1992).

\bibitem{art}
J.~Dalibard, F.~Gerbier, G.~Juzeli\~unas, and P.~\"Ohberg, \refttl{Colloquium: Artificial gauge potentials for neutral atoms}
{\sl Rev.\ Mod.\ Phys} {\bf 83} 1523 (2011);
F.~Wilczek and A.~Zee, \refttl{Appearance of gauge structure in simple dynamical systems}{\sl Phys.\ Rev.\ Lett.} {\bf 52} 2111 (1984);
M.V.~Kazan, \refttl{Analog of the Aharonov--Bohm effect in superfluid He${}^3$-A}{\sl Pis'ma Zh.\ Eksp.\ Theor.\ Phys} {\bf 41}, No.\ 9,
396-398 (1985);
J.~Moody and F.~Wilczek, \refttl{Realizations of magnetic-monopole gauge fields: diatoms and spin precession}{\sl Phys.\ Rev.\ Lett.} {\bf 56} 893 (1986).

\bibitem{vort}
L.M.~Pismen, {\sl Vortices in nonlinear fields}, Clarendon Press, Oxford, 1999 (esp.\ Ch.\ 4 and references therein);
N.B.~Kopnin, {\sl Theory of nonequilibrium superconductivity}, Oxford University Press, Oxford, 2001.

\bibitem{ior}
S.V.~Iordansky, \refttl{On the mutual friction between the normal and superfluid components}{\sl Ann.\ Phys. (NY)} {\bf 29} (1964) 335--349;
E.~Sonin, \refttl{Magnus force in superfluids and superconductors}{\sl Phys.\ Rev.} {\bf B55} (1997) 485;
M.~Stone, \refttl{Iordanskii force and the gravitational Aharonov--Bohm effect for a moving vortex}{\sl Phys.\ Rev.} {\bf B61} 11780-11786 (2000).

\bibitem{rel}
A.~Vilenkin and E.P.S.~Shellard, {\sl Cosmic strings and other topological defects}, Cambridge University Press, Cambridge, 1994
(esp.\ Ch.\ 8);
M.G.~Ahlford, J.~March-Russell, and F.~Wilczek, \refttl{Enhanced baryon number violation due to cosmic strings}{\sl Nucl.\ Phys.} {\bf B328} (1989) 140-158;
A.C.~Davis and A.P.~Martin, \refttl{Global strings and the aharonov--Bohm effect}{\sl Nucl.\ Phys.} {\bf B419} (1994) 341-351;
A.C.~Davis, A.P.~Martin, and N.~Ganoulis, \refttl{Charged-particle scattering from electroweak and semi-local strings}{\sl Nucl.\ Phys.} {\bf B419} (1994) 323-340.

\bibitem{FLRv} P\'eter Forg\'acs, \'Arp\'ad Luk\'acs, and Tomasz Roma\'nczukiewicz, {\sl in preparation}.

\bibitem{FLR} P\'eter Forg\'acs, \'Arp\'ad Luk\'acs, and Tomasz Roma\'nczukiewicz, \refttl{Negative radiation pressure exerted on kinks}
{\sl Phys.\ Rev.} {\bf D77}, 125012 (2008).

\bibitem{trom} T.~Roma\'nczukiewicz, \refttl{Negative radiation pressure in case of two interacting fields}{\sl Acta.\ Phys.\ Polonica} {\bf B39} (2008) 3449-3462.

\bibitem{sukhov} S.~Sukhov and D.~Dogariu, \refttl{Negative nonconservative forces: optical “tractor beams” for arbitrary objects}
{\sl Phys.\ Rev.\ Lett.} {\bf 107}, 203602 (2011);
A.~Novitsky, C.-W.~Qiu, and H.~Wang, \refttl{Single gradientless light beam drags particles as tractor beams}
{\sl Phys.\ Rev.\ Lett.} {\bf 107}, 203601 (2011).

\bibitem{AB} Y.~Aharonov and D.~Bohm, \refttl{Significance of Electromagnetic Potentials in Quantum Theory}{\sl Phys. Rev.} {\bf 115} (1959) 485.

\bibitem{ABf}
S.~Olariu and I.I.~Popescu, \refttl{Quantum effects of electromagnetic fluxes}{\sl Rev.\ Mod.\ Phys.} {\bf 57} 339-436 (1985);
A.L.~Shelankov, \refttl{Magnetic force exerted by the Aharonov--Bohm line}{\sl Europhys.\ Lett.} {\bf 43} (6) pp.\ 623-628 (1998).

\bibitem{AbrSteg} M.~Abramowitz and I.A.~Stegun (eds.), {\sl Handbook of mathematical functions}, National Bureau of Standards, 1964.
\end{thebibliography}
\end{document}